\documentclass[prb,preprint,superscriptaddress,showpacs,showkeys,preprintnumbers,amsmath,amssymb]{revtex4-1}
\usepackage{graphicx,color}

\begin{document}

\pacs{71.15.Mb, 75.70.Cn, 85.75.-d}
\keywords{Heusler alloy, half metal, heterostructure, density functional theory} 

\title{The fate of half-metallicity near interfaces: the case of NiMnSb/MgO and NiMnSi/MgO}

\author{Rui-Jing \surname{Zhang}}
\email{ruijing.struckmeier@physik.uni-augsburg.de}
\author{Ulrich \surname{Eckern}}
\affiliation{Institute of Physics, University of Augsburg, 86135 Augsburg, Germany}
\author{Udo \surname{Schwingenschl\"ogl}}
\email{udo.schwingenschlogl@kaust.edu.sa}
\affiliation{KAUST, PSE Division, Thuwal 23955-6900, Kingdom of Saudi Arabia}

\begin{abstract}
The electronic and magnetic properties of the interfaces between the half-metallic Heusler
alloys NiMnSb, NiMnSi and MgO have been investigated using first-principles density-functional 
calculations with projector augmented wave potentials generated in the generalized gradient
approximation. In the case of the NiMnSb/MgO (100) interface the half-metallicity is lost,
whereas the MnSb/MgO contact in the NiMnSb/MgO (100) interface maintains a substantial degree
of spin polarization at the Fermi level ($\sim 60$\%). Remarkably, the NiMnSi/MgO (111) interface
shows 100\% spin polarization at the Fermi level, despite considerable distortions at the
interface, as well as rather short Si/O bonds after full structural optimization. This behavior
markedly distinguishes NiMnSi/MgO (111) from the corresponding NiMnSb/CdS and NiMnSb/InP interfaces.
\end{abstract}

\maketitle

\section{Introduction}

Ferromagnetic half-metals such as CrO$_2$ and Heusler alloys are key materials for
technological applications, e.g., spin-injection devices, spin filters, tunnel junctions
and giant magnetoresistance devices. Heusler alloys have various advantages that make them
attractive for applications. One is the relatively high Curie temperature as compared with other
half-metallic systems. The other is the structural similarity to the zinc-blende structure,
which is adopted by binary semiconductors widely used in technology. In particular, the NiMn$Z$
($Z$ = Si, P, Ge, As) family of compounds seems to be promising in this context. \cite{dinh08}
For example,
NiMnSi has been predicted to have a Curie temperature of 1050~K, \cite{katayama07} exceeding
the 730~K of NiMnSb, \cite{webster88} which calls for a detailed study of the electronic and
magnetic properties of heterostructures containing this compound.

The functionality of nanoscale devices depends crucially on the transport across the
interfaces between the different components. Thus, the electronic and magnetic properties
at interfaces have attracted huge interest in recent years. It has been shown by theoretical
calculations as well as experiments that the half-metallicity of Heusler alloys is lost at
surfaces and interfaces due to symmetry breaking.
\cite{lezaic05,galanakis02,galanakis05a,jenkins04,jenkins01,wijs01,galanakis05b,debernardi05,lezaic06}
However, it has also been argued that the half-metallicity is conserved when the S atoms
sit exactly on top of the Sb atoms at the NiMnSb/CdS (111) contact. \cite{wijs01} The same
phenomenon appears at the NiMnSb/InP (111) contact when the P atoms sit exactly on top of the
Sb atoms. \cite{galanakis05b} Both cases are interfaces between NiMnSb and a semiconductor and
have the anion-anion bonds at the interface. In fact, the (111) interfaces are special in this
respect, and thus generally promising to conserve half-metallicity, whereas in (100) interfaces,
for example, the anions are coordinated by a mixture of main-group and transition metals.

We will show that half-metallicity is indeed preserved at the (111) interface between the
half-Heusler alloy NiMnSi and the insulator MgO.
In the following, two orientations of the NiMnSb(Si)/MgO interface are investigated:
the (100) interface as it is most commonly applied in experiments, and the (111)
interface as it contains only one component in each plane. 

\section{Structure and computational method}

Calculations have been performed for supercells consisting of NiMnSb
or NiMnSi and MgO, using the projector augmented wave method \cite{blochl94} within the
generalized gradient approximation \cite{perdew96} and density-functional theory.
\cite{kohn64} In order to take
into account all degrees of freedom for the NiMnSb/MgO (100), NiMnSb/MgO (111) and
NiMnSi/MgO (111) interfaces, we have conducted full structural optimizations for the cell
parameters and internal coordinates. A plane-wave basis set with 450 eV energy
cutoff is used; the convergence criterion of the total energy change is set to 10$^{-4}$ eV.
The $k$-point mesh is taken as $17\times17\times1$ for the (100) interfaces and
$7\times7\times1$ for the (111) interfaces. All technical parameters have been tested
carefully to ensure accurate results. For the calculations we employ the Vienna ab-initio
simulation package (VASP). \cite{vasp}

The crystal structure of the half-Heusler alloys NiMnSb and NiMnSi is face centered cubic
($F{\overline 4}$3m), consisting of four sublattices: Ni at the
(0,0,0), Mn at the (1/4,1/4,1/4), Sb or Si at the (3/4,3/4,3/4) and vacancies at
the (1/2,1/2,1/2) sites. The calculated lattice constant of NiMnSb (5.91~\AA) is
within 1\% accuracy to the experimental value (5.93~\AA), \cite{webster88} whereas
NiMnSi has not yet been grown experimentally. However, a lattice constant
of 5.4~\AA\ has been theoretically obtained in Ref.\ \onlinecite{galanakis08}, which agrees well
with our value of 5.37~\AA. The [100] direction of NiMnSb and [110] direction of MgO show
a nearly perfect lattice match, whereas the lattice mismatch between NiMnSi (100) and MgO (100)
is about 10\%.
This explains why the NiMnSb/MgO interface
could be realized experimentally, \cite{turban02,turban02b,sicot06,tsunegi09}
but not the NiMnSi/MgO (100) system.

\begin{figure}
\includegraphics[width=0.4\textwidth,clip]{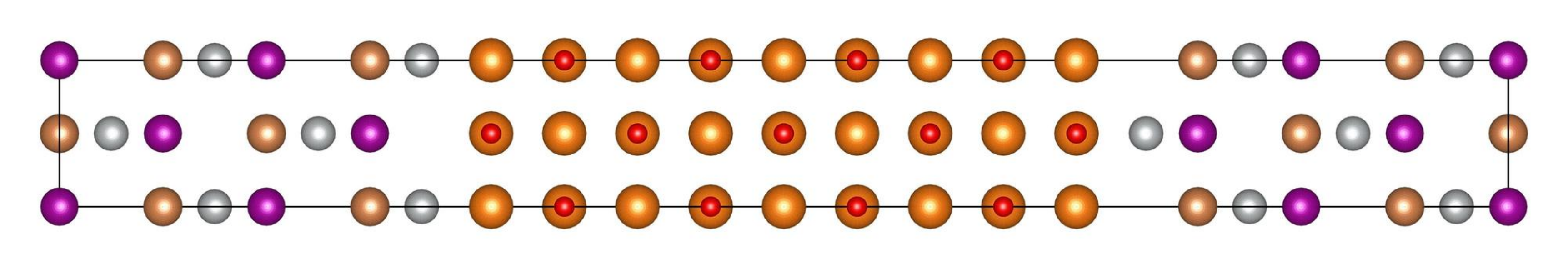}(a)\\
\includegraphics[width=0.4\textwidth,clip]{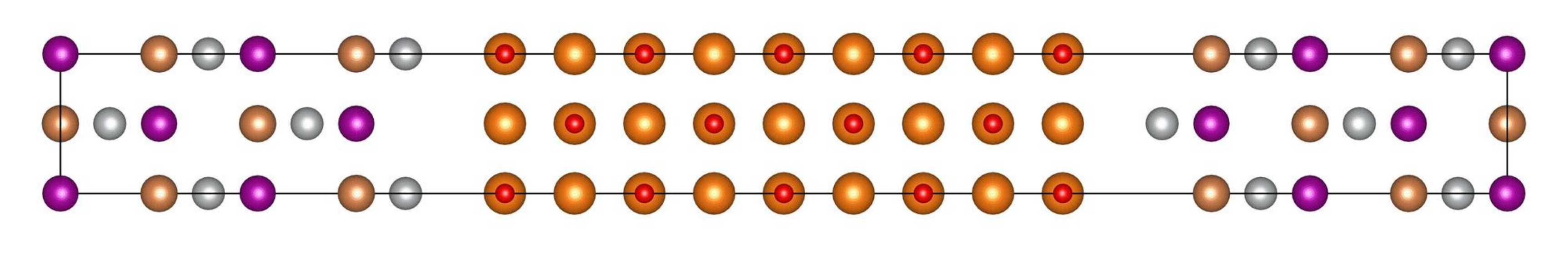}(b)\\
\includegraphics[width=0.4\textwidth,clip]{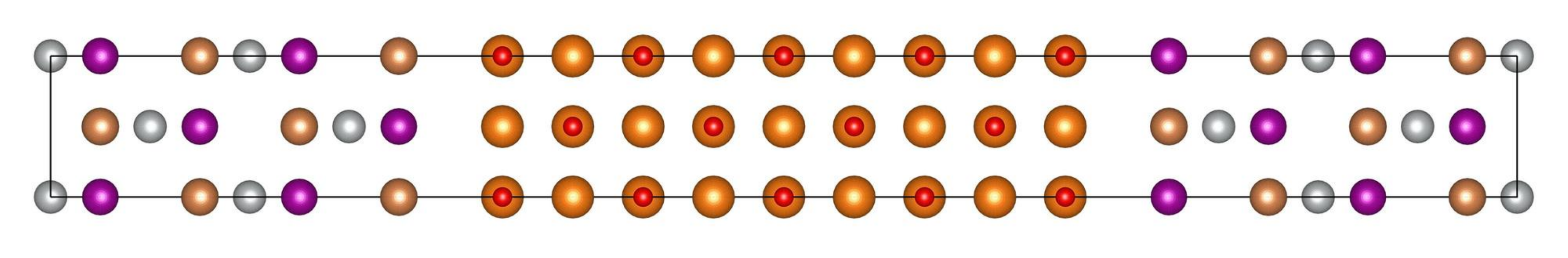}(c)\\
\includegraphics[width=0.4\textwidth,clip]{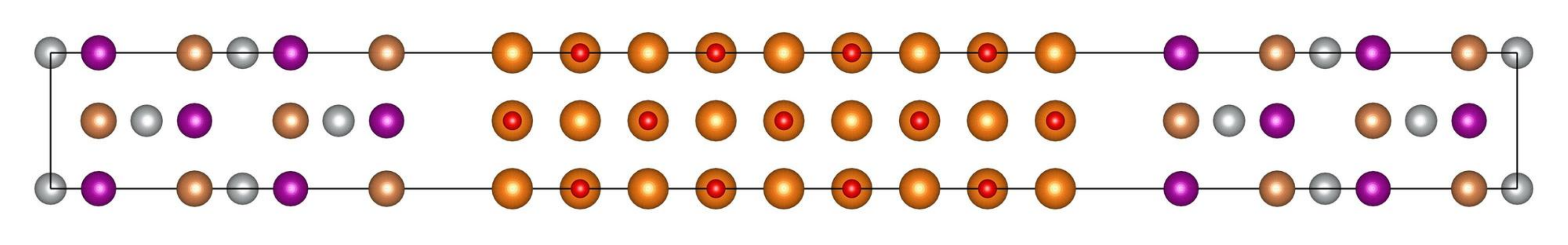}(d)
\caption{Side views of the supercells used for modeling the NiMnSb/MgO/NiMnSb (100)
heterostructure: (a) Ni/OMg-terminated interface, (b) Ni/MgO-terminated interface,
(c) MnSb/OMg-terminated interface, and (d) MnSb/MgO-terminated interface. Color code:
Ni (grey), Mn (purple), Sb (bronze), Mg (orange), O (red). The interface areas are
between 17.66 \AA$^2$ (a) and 17.44 \AA$^2$ (d).}
\label{structure-100}
\end{figure}

For the (100) interface we employ multiple supercells with 15 layers of NiMnSb and nine layers
of MgO. This thickness is sufficient for the central regions of both the half-metal
and the MgO to exhibit their bulk properties. As is shown in Fig.~\ref{structure-100}, by
building four supercells we construct all possible interfaces with four different terminations,
where the two interfaces within one supercell possess the same configuration.
For the (111) interface the supercells consist of 25 layers of NiMnSb or NiMnSi
and 13 layers of MgO with Sb/O or Si/O contacts. More structural details will be elaborated
in the following section.

\begin{table}
\caption{Bond lengths at the NiMnSb/MgO (100) interfaces.}
\smallskip
\begin{tabular}{l|c|c}\hline
Interface termination & Bond type & Bond length (\AA)\\\hline\hline
Ni/OMg               & Ni-O      & 2.0 \\ \hline
Ni/MgO               & Ni-Mg     & 3.1 \\ \hline
                     & Mn-O      & 2.5 \\ \cline{2-3}
\raisebox{1.5ex}[-1.5ex]{MnSb/OMg} & Sb-O      & 3.0 \\ \hline
                     & Mn-Mg     & 3.9 \\ \cline{2-3} 
\raisebox{1.5ex}[-1.5ex]{MnSb/MgO} & Sb-Mg     & 3.8 \\ \hline
\end{tabular}
\label{bonds-100}
\end{table}

\section{Results and discussion}

\subsection{NiMnSb/MgO (100) interfaces}

Equilibrium structures for the four supercells in Fig.~\ref{structure-100}
have been obtained by full structural
relaxation. The bond types and lengths at the different interfaces are summarized in
Table~\ref{bonds-100}.
For instance, Ni/OMg termination means that a layer containing only Ni atoms and a
layer containing both Mg and O atoms compose the interface with direct Ni-O bonds. The length
of these bonds is found to be the shortest at the Ni/OMg-terminated interface (2~\AA).
The optimized Ni-Mg bond length is 3.1~\AA\ at the interface in the case of Ni/MgO termination.
We note that the MnSb/OMg-terminated interfaces show a stronger reconstruction with
bond lengths between Mn-O (2.5~\AA) and Sb-O (3.0~\AA). The Mn-Mg (3.9~\AA) and
Sb-Mg (3.8~\AA) bonds are the longest at the MnSb/MgO-terminated interface. (We
only consider the nearest-neighbor atoms at the interface.) It is
understandable that the interface thickness is larger for the case of metal
(Ni or MnSb)/MgO termination than for metal/OMg termination, because of the repulsion
between the metal and Mg atoms. 

\begin{table}
\caption{Work of separation $W$ for the NiMnSb/MgO (100) interface.}
\smallskip
\begin{tabular}{l|c}\hline
Interface termination & $W$ (eV) \\\hline\hline
Ni/OMg               & $+ 0.532$  \\ \hline
Ni/MgO               & $- 0.410$ \\ \hline
MnSb/OMg             & $+ 1.843$ \\ \hline
MnSb/MgO             & $- 0.467$ \\ \hline
\end{tabular}
\label{work-100}
\end{table}

\begin{figure*}
\includegraphics[width=0.4\textwidth,clip]{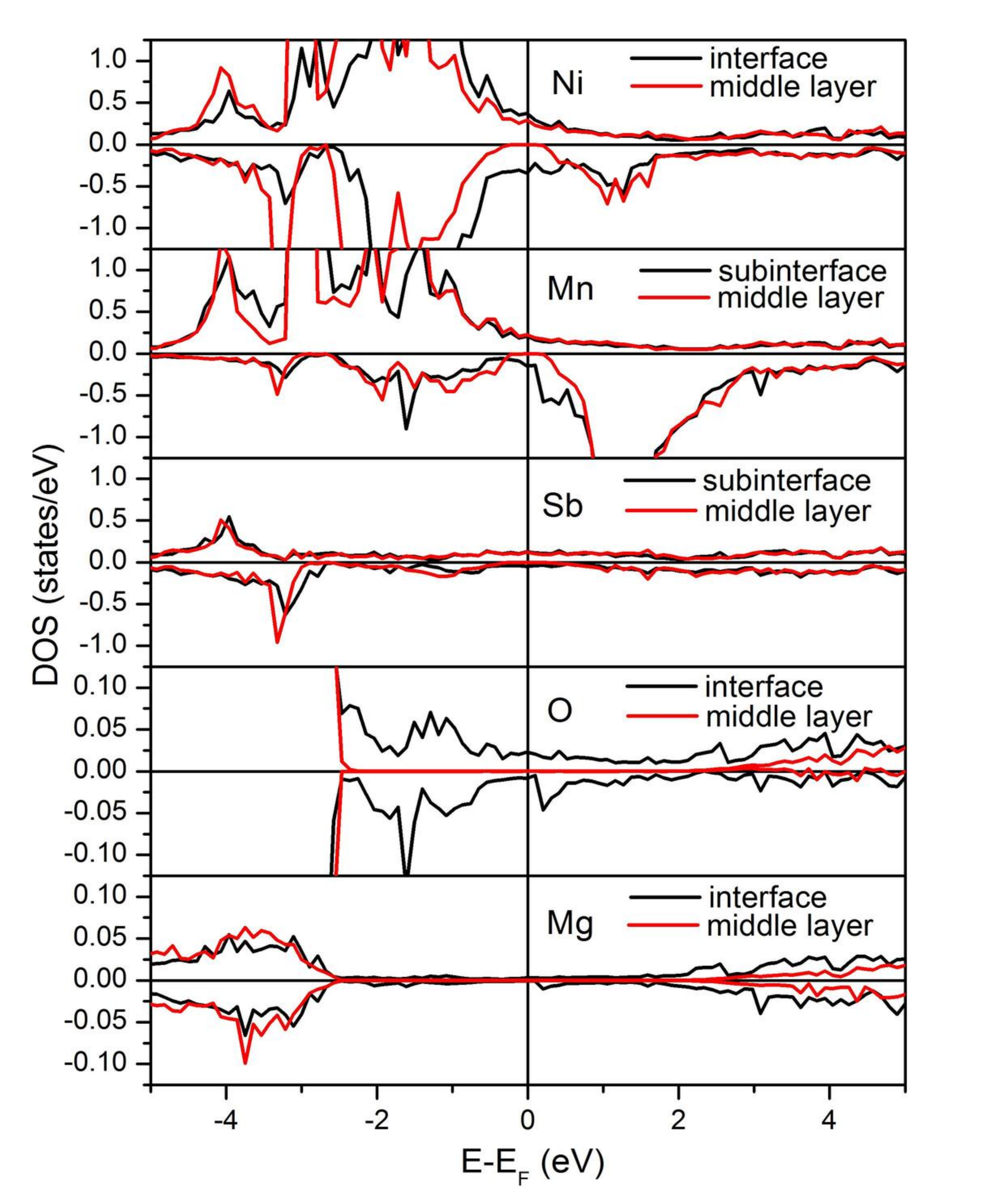}(a)
\includegraphics[width=0.4\textwidth,clip]{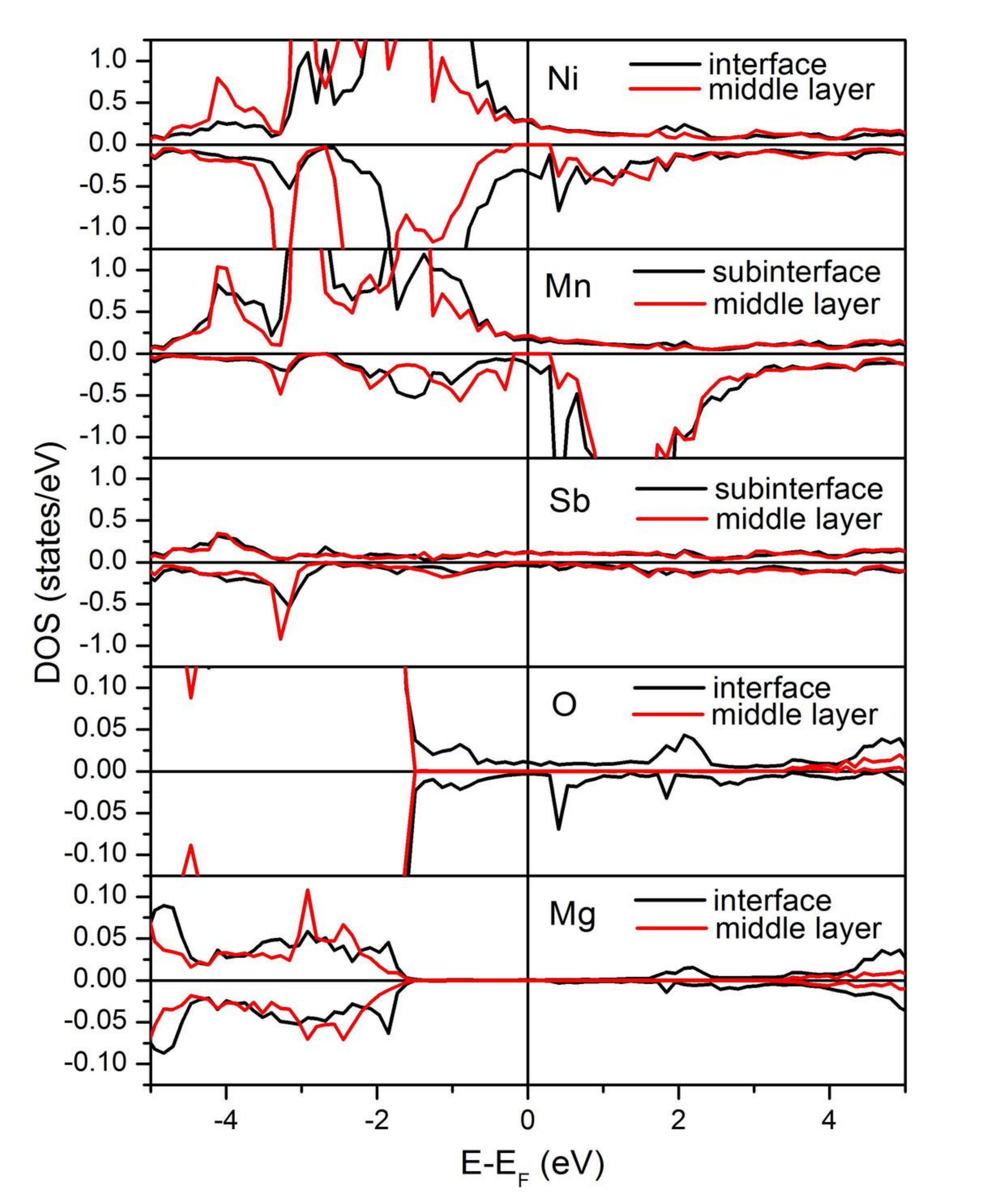}(b)
\includegraphics[width=0.4\textwidth,clip]{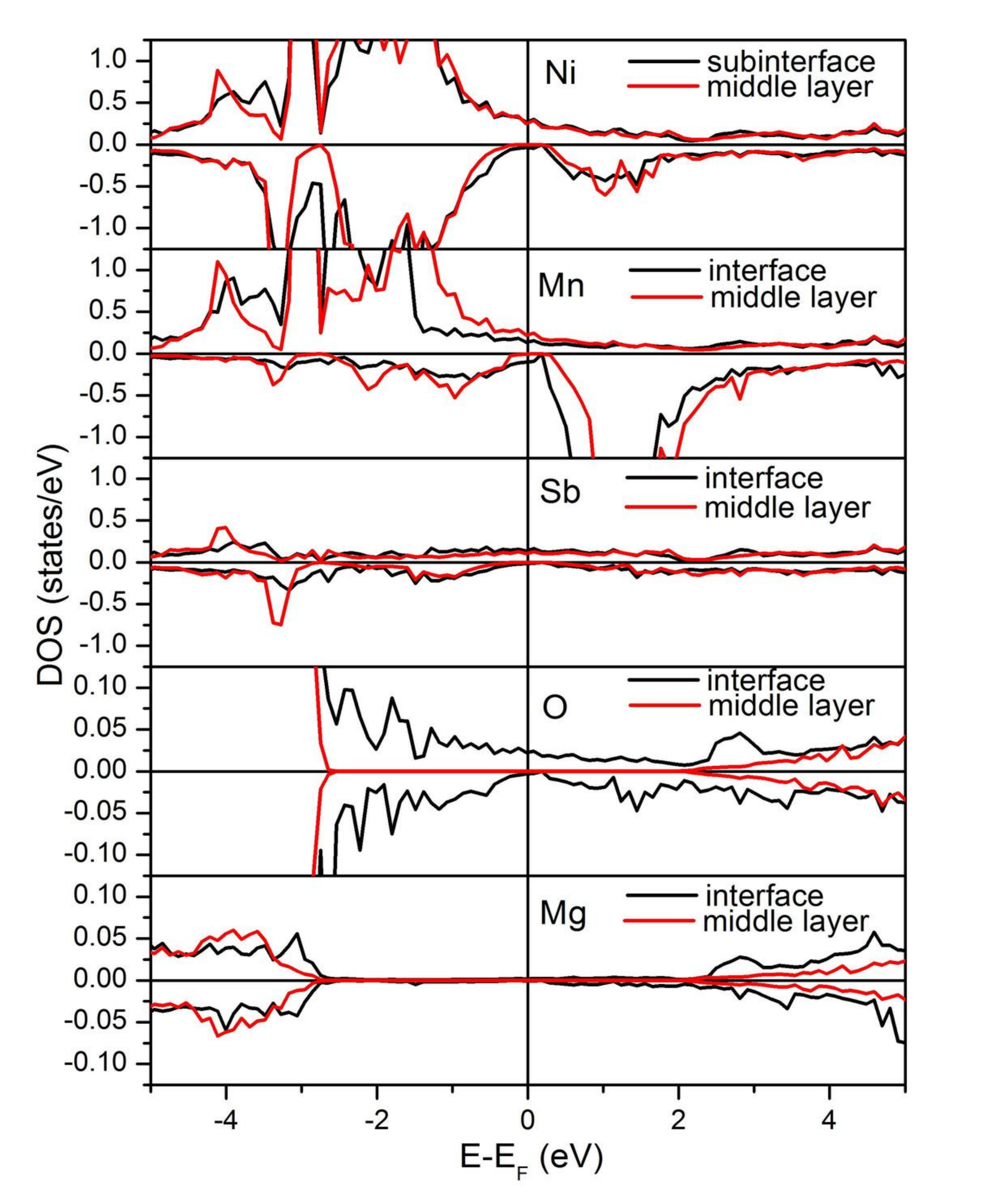}(c)
\includegraphics[width=0.4\textwidth,clip]{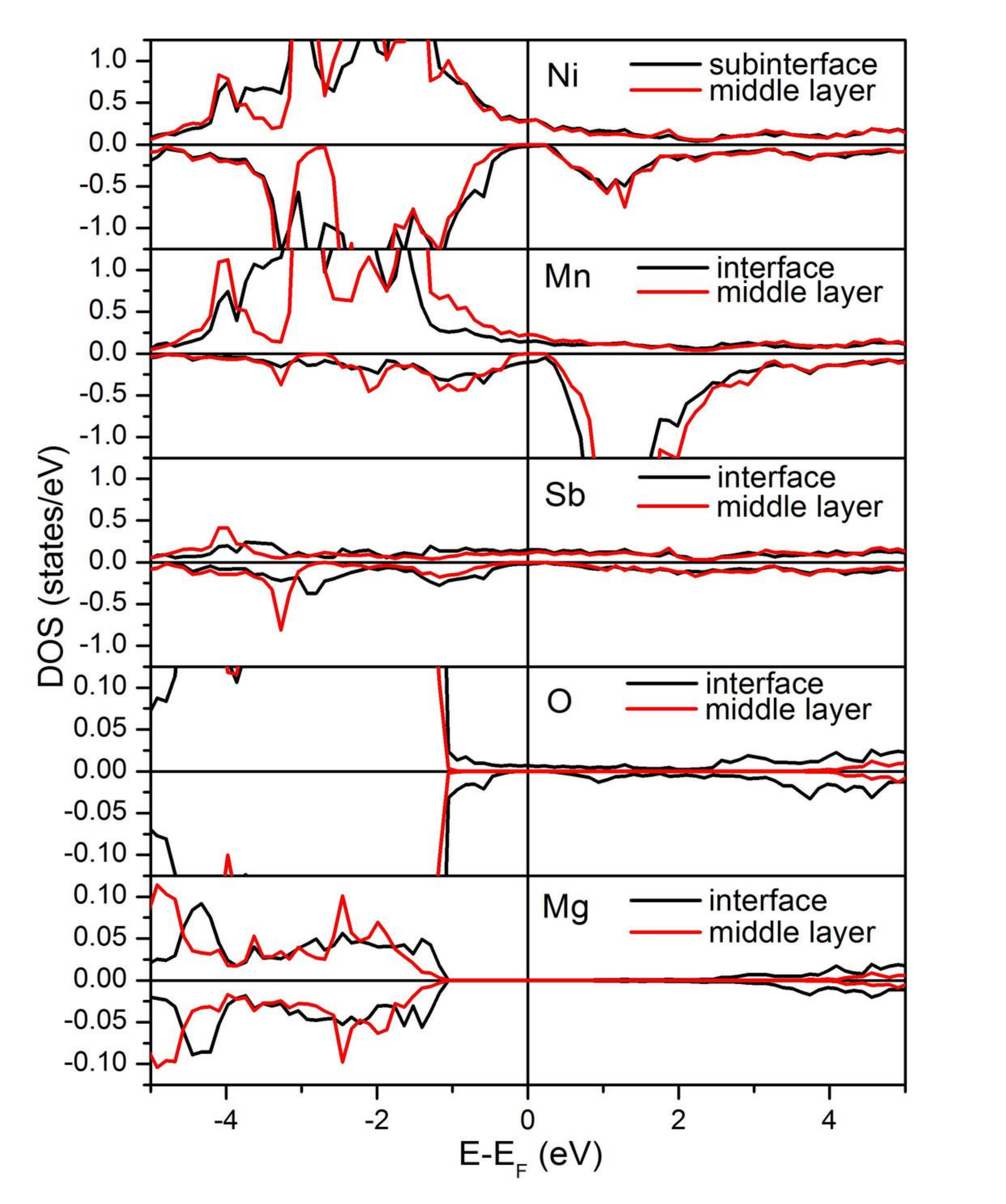}(d)
\caption{Atom- and spin-resolved DOS for the
atoms at the interface and in the middle layer of the NiMnSb or MgO slab
of NiMnSb/MgO (100): (a) Ni/OMg-terminated interface,
(b) Ni/MgO-terminated interface, (c) MnSb/OMg-terminated interface, and
(d) MnSb/MgO-terminated interface.}
\label{dos-100}
\end{figure*}

The bonding strength of the interfaces can be characterized by the work of separation,
\begin{equation}
W = \frac{1}{2} \left[ E_{\rm NiMnSb \; slab} +  E_{\rm MgO \; slab} - E_{\rm NiMnSb/MgO} \right],
\end{equation}
where $E_{\rm NiMnSb \; slab}$ and $E_{\rm MgO \; slab}$ are
the total energies of the isolated slabs (surrounded
by vacuum within a supercell, where the lattice parameters are kept the same as obtained
for the combined system in equilibrium, and no further structure relaxation is performed),
and $E_{\rm NiMnSb/MgO}$ is the total energy of the heterostructure.
The factor one half represents the two equivalent interfaces in the supercell.
Results obtained for the work of separation are listed in Table~\ref{work-100}. 

We find that the interfaces with Ni-Mg, Mn-Mg, and Sb-Mg bonds exhibit negative work
of separation, which indicates that metal/MgO-terminated interfaces are difficult to be
formed as cation-cation interaction is energetically unstable. On the contrary, the
Ni/OMg- and MnSb/OMg-terminated interfaces show positive work of separation, with a
three times larger value for the latter, which implies that Mn-O and Sb-O bonds are
energetically favored over Ni-O bonds. Previous investigations have predicted 100\% spin
polarization for the NiMnSb/CdS (111) and the NiMnSb/InP (111) interface.
\cite{wijs01,galanakis05b} Both cases are characterized by a long anion-anion bond,
but the bonding strengths of the interfaces have not been evaluated. 

Figure \ref{dos-100} shows the projected density of states (DOS; per atom here and in
the following) for atoms at the four NiMnSb/MgO interfaces.
The interface and sub-interface layers of the NiMnSb region and the interface
layer of the MgO region are taken into consideration. It can be seen that the bulk
properties are well reproduced in the middle layers of both NiMnSb and MgO. However,
the half-metallicity completely vanishes at the Ni/OMg- and Ni/MgO-terminated interfaces,
where there are one Ni and two O atoms at the interface. The reason is that not only
Ni (due to the short Ni-O bond) but also the Mn atom in the sub-interface layer
is oxidized (due to the excess O atom). For the MnSb/OMg- and
MnSb/MgO-terminated interfaces the interaction between Ni in the sub-interface layer
and O at the interface is screened by the MnSb layer in between. The Sb and O atoms
exhibit a weak interaction since both are anions.

As a consequence, Ni in the sub-interface
layer and Sb in the interface layer preserve almost the half-metallic state with
a nearly zero DOS for the spin minority band and a non-zero DOS for the spin majority
band at the Fermi level. Mn is less oxidized due to a longer Mn-O bond as compared
with Ni at the Ni/OMg- and Ni/MgO-terminated interfaces. All this contributes to a
higher spin polarization for the MnSb-terminated than for the Ni-terminated
interface, as shown in Table~\ref{ratio-100}. For all possible configurations in Fig.~\ref{dos-100},
the Mn atom is strongly oxidized at the interface, which reduces the spin polarization.
Moreover, the O atom of MgO at the interface contributes some states at the Fermi
level. Especially at the metal/OMg-terminated interfaces, the direct bonding
(Ni-O, Mn-O or Sb-O) results in stronger interaction and enhanced charge transfer
between the Heusler alloy and O. 

From Fig.~\ref{dos-100} we obtain the spin polarization ratio $P$ at the Fermi level for
each interface (see Table~\ref{ratio-100}) as
\begin{equation}
P = \frac{N_\uparrow (E_F) - N_\downarrow (E_F)}{N_\uparrow (E_F) + N_\downarrow (E_F)}
\end{equation}
in terms of the spin-resolved DOS, $N_\uparrow (E_F)$ and $N_\downarrow (E_F)$.
The spin polarization is very low for the
Ni-terminated interface and high for the MnSb-terminated interface (up to around 60\%).
The spin magnetic moments at the interface and for the most bulk-like atoms are
listed in Table~\ref{moments-100}. In the case of the MnSb-terminated interface the Mn atom directly
interacts with MgO as well as the Ni and Sb atoms. Consequently, the charge transfer
between these atoms brings additional electrons into the Mn majority band. We observe
a gain of 0.25$\mu_B$ as compared with the bulk-like value, where $\mu_B$ is the Bohr
magneton, while there is only a 0.1$\mu_B$ gain in the case of Ni-termination.

\begin{table}
\caption{Spin polarization ratio $P$ at the (100) interfaces for (a) the first
two layers of NiMnSb, and (b) the first two layers of
NiMnSb and the first layer of MgO. The numbers in parentheses are
$N_\uparrow (E_F)$ and $N_\downarrow (E_F)$, respectively, in units of 1/eV.}
\smallskip
\begin{tabular}{c||c|c|c|c}\hline
         & Ni/OMg & Ni/MgO & MnSb/OMg & MnSb/MgO\\\hline\hline
         & 13\% &  6\% & 56\% & 60\% \\                
\raisebox{1.5ex}[-1.5ex]{(a)} & (0.686, 0.532) & (0.566, 0.498) & (0.599, 0.156) & (0.576, 0.143) \\ \hline
            & 15\% &  8\% & 58\% & 61\% \\
\raisebox{1.5ex}[-1.5ex]{(b)} & (0.738, 0.550) & (0.589, 0.506) & (0.609, 0.163)  & (0.590, 0.143) \\ \hline
\end{tabular}
\label{ratio-100}
\end{table}

\begin{table}
\caption{Spin magnetic moments (per atom, in $\mu_B$) at the (100) interfaces
for the first two layers of NiMnSb and the first layer of MgO at the interface,
for the four terminations. An asterisk denotes atoms in the sub-interface layer.}
\smallskip
\begin{tabular}{ll||c|c|c|c|c|c}\hline
          & & Ni & Mn & Sb & O & Mg & total\\\hline\hline
 Ni/OMg & Interface & 0.28 & 3.78* & $-$0.05 & 0.01 & 0.00 & 4.03 \\
        & Bulk-like & 0.25 & 3.67 & $-$0.06 & 0.00 & 0.00 & 3.86 \\ \hline
 Ni/MgO & Interface & 0.37 & 3.76* & $-$0.05* & 0.01 & 0.00 & 4.10 \\
        & Bulk-like & 0.25 & 3.68 & $-$0.06 & 0.00 & 0.00 & 3.87 \\ \hline
 MnSb/OMg \; & Interface & 0.20* & 3.91 & $-$0.07 & 0.01 & 0.00 & 4.05 \\
        & Bulk-like & 0.24 & 3.67 & $-$0.06 & 0.00 & 0.00 & 3.85 \\ \hline
 MnSb/MgO \; & Interface & 0.23* & 3.91 & $-$0.09 & 0.00 & 0.00 & 4.05 \\
        & Bulk-like & 0.24 & 3.68 & $-$0.06 & 0.00 & 0.00 & 3.86 \\ \hline
\end{tabular}
\label{moments-100}
\end{table}

\subsection{NiMnSb/MgO and NiMnSi/MgO (111) interfaces}

\begin{figure*}
\includegraphics[width=0.5\textwidth,clip]{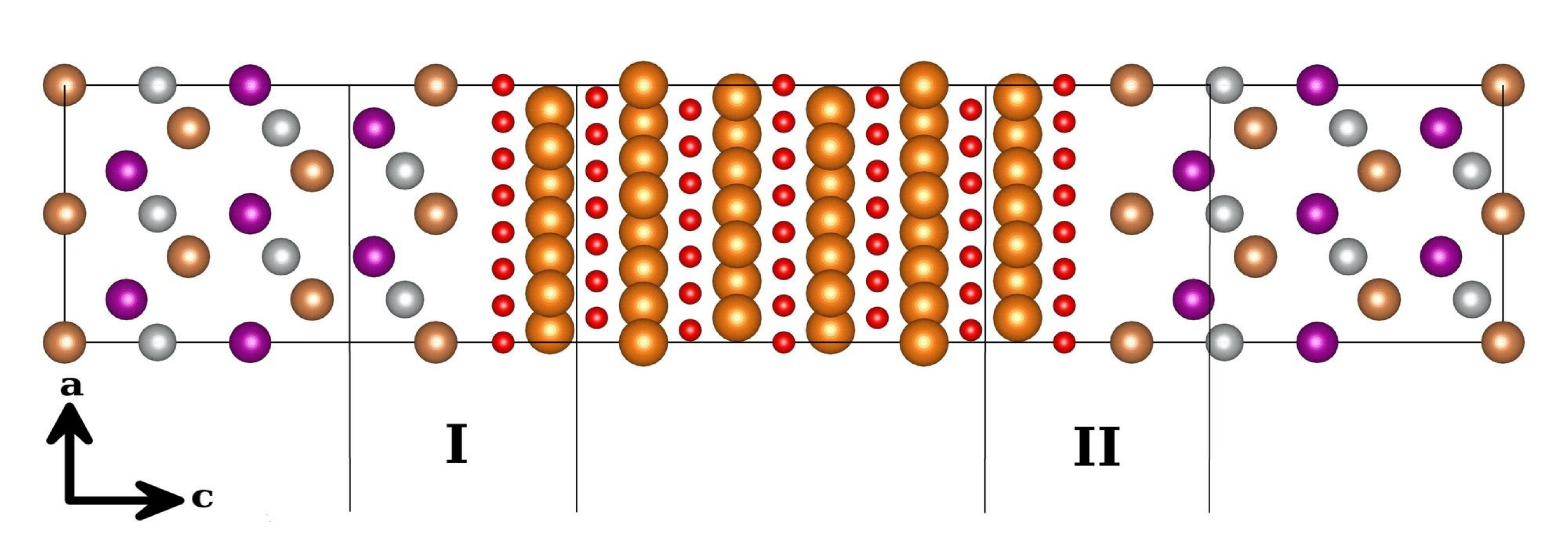}\hspace{2em}
\includegraphics[width=0.25\textwidth,clip]{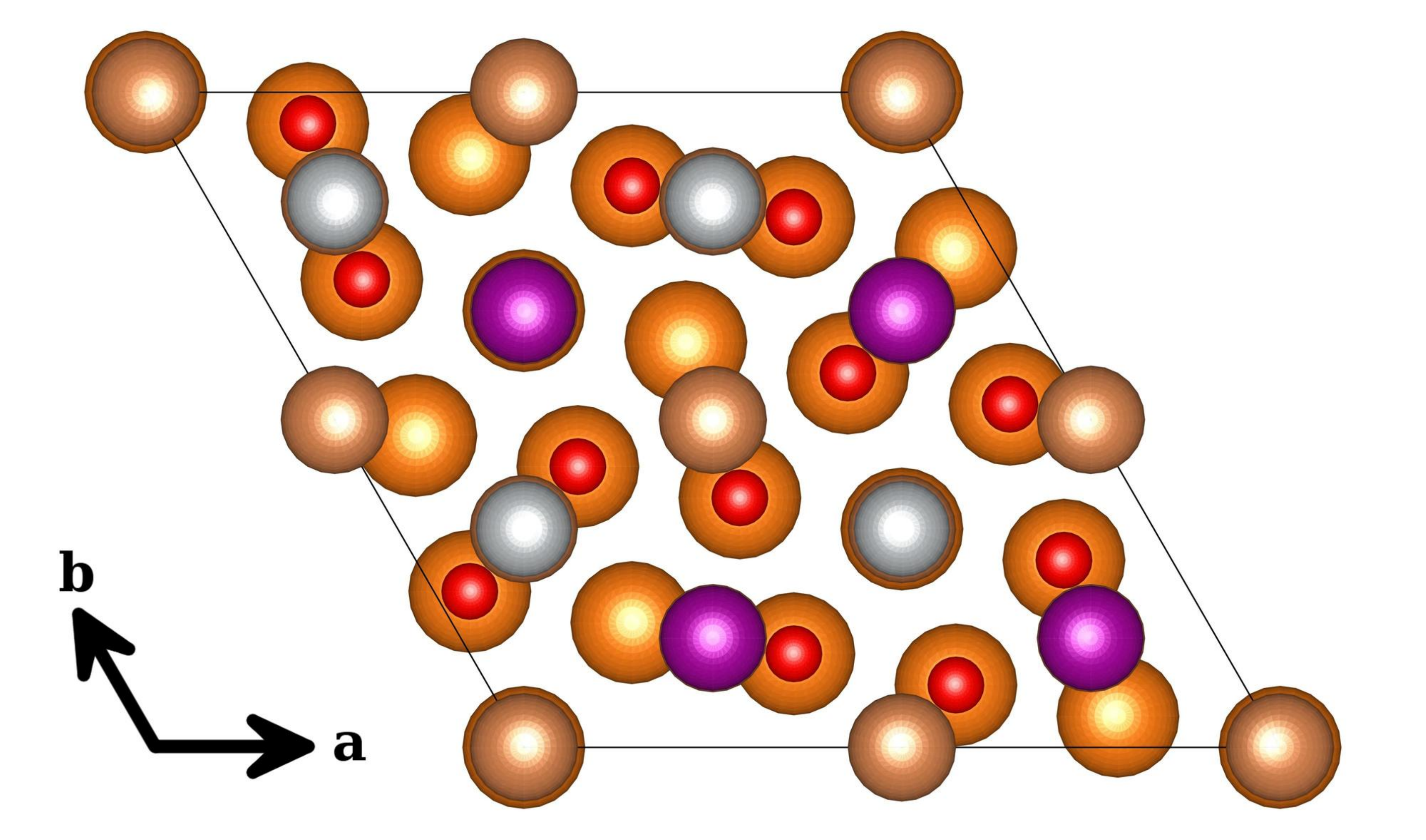}
\caption{View along the $b$-axis (left) and $c$-axis (right) of the NiMnSb/MgO/NiMnSb
(111) heterostructure with Sb/O-terminated interface. The supercell contains
two interfaces: I (left, three layers of NiMnSb and two layers of MgO,
from left to right) and II (right, vice versa). By substituting Sb with Si we obtain
the NiMnSi/MgO system. The color code is the same as in Fig.~\ref{structure-100}.
Note that the supercell is chosen such that the structures of the interfaces I and II
are as close as possible to each other. The interface area is 53.77 \AA$^2$.}
\label{structure-111}
\end{figure*}

Figure \ref{structure-111} shows the supercell of the NiMnSb/MgO (111) (or NiMnSi/MgO (111))
interface with Sb/O-(or Si/O-)termination. The stacking of layers at interface I
from right to left is Mg-O-Sb(or Si)-Ni-Mn, and at interface II
from left to right Mg-O-Sb(or Si)-vacancies-Mn. The Mn atoms are placed at maximal
distance to the O atoms, because Mn is most likely to be oxidized at the interface,
which is not favorable to maintain half-metallicity.
By building a hexagonal supercell with $a = 7.9$~\AA\ and 167 atoms,
the minimum lattice mismatch for the (111) plane is found to be 6\% between NiMnSb and
MgO, and 3.7\% between NiMnSi and MgO. The in-plane lattice
constant is initially set to the value of MgO (7.9~\AA), because in experiments the
Heusler alloy is grown on the MgO substrate. Therefore, the NiMnSb (or NiMnSi)
lattice constant should change to match the MgO lattice. The layers of the Heusler
alloys and MgO are kept initially bulk-like, e.g., approximately
0.8~\AA\ apart for NiMnSi and 1.2~\AA\ apart for MgO. 

The projected DOS demonstrates that at the NiMnSb/MgO (111) interface the spin polarization
ratio at the Fermi level is very low. However, the energy gap in the minority
band, which is shifted above the Fermi level for interface I and below the Fermi level
for interface II, still exists. This indicates that spin polarization at the
Fermi level can be recovered by proper doping with NiMnSb.

\begin{figure}
\includegraphics[width=0.5\textwidth,clip]{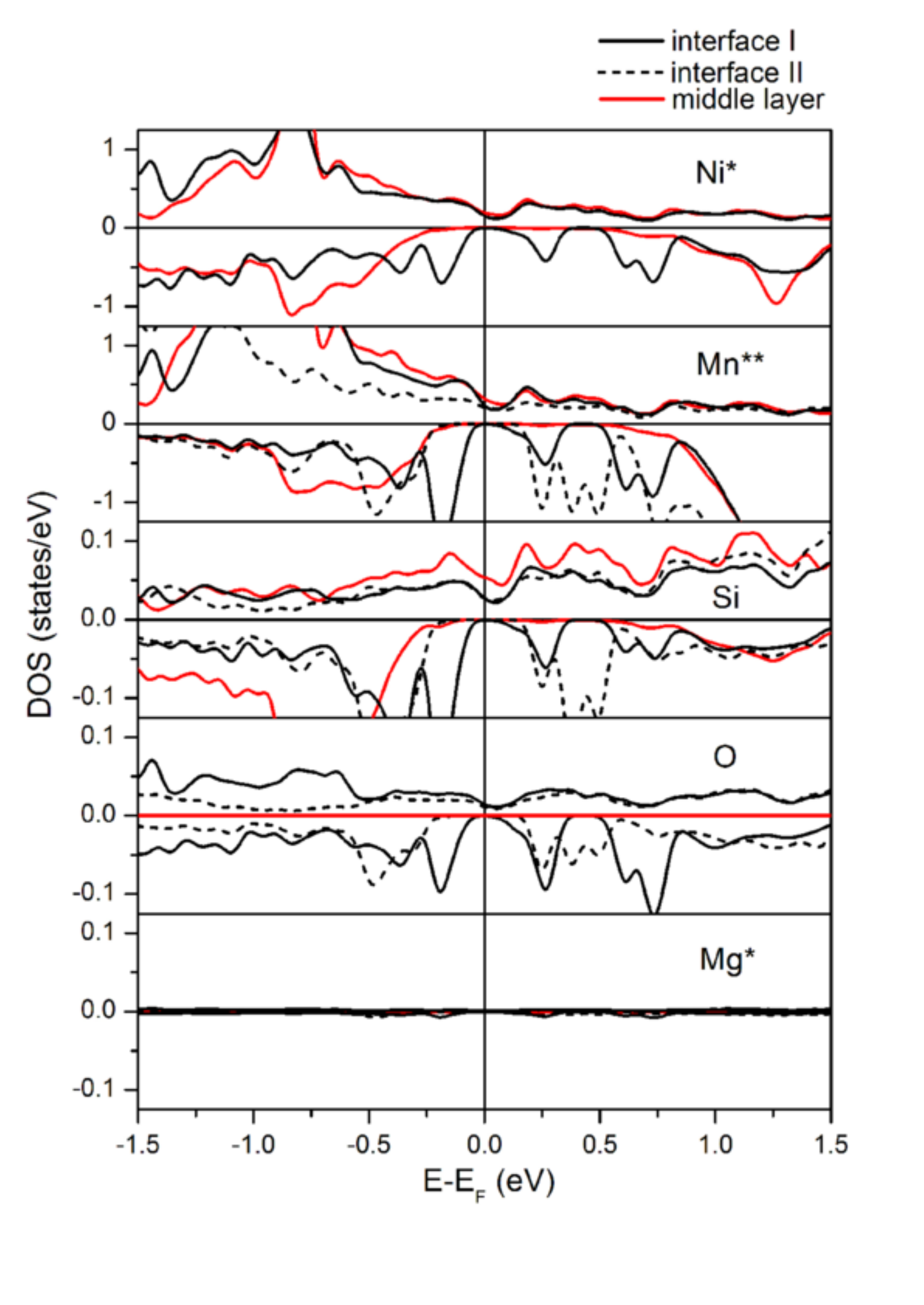}
\caption{Atom- and spin-resolved DOS for the atoms at the interface and in the middle
layer of the NiMnSi or MgO slab of NiMnSi/MgO (111) with Si/O termination. Single and double
asterisks denote atoms in the sub-interface and sub-sub-interface layers, respectively.}
\label{dos-111}
\end{figure}

Turning to the NiMnSi/MgO
(111) system, Fig.~\ref{dos-111} shows the projected DOS for atoms at the interface,
including the first three layers of NiMnSi and first two layers of MgO. The Fermi level lies
in the gap of the minority states, which reflects the conservation of half-metallicity.
Both interfaces show 100\% spin polarization, though interface II with a vacancy
layer between Si and Mn has a much larger band gap in the minority channel for
all atoms at the interface than interface I with three metal layers.

\begin{table}
\caption{Detailed interface configuration, spin polarization, and work of
separation for the NIMnSi/MgO (111) interface, where the arrow in the order of stacking
layers indicates `from right to left' for interface I and `from left to right' for
interface II. $W_{\rm I+II}$ is the aggregate work of separation for interfaces I and II.
(vac = vacancies)}
\smallskip
\begin{tabular}{c||c|c|c|c}\hline
          & Order of              & Average bond         & Spin         & $W_{\rm I+II}$ \\
\raisebox{1.5ex}[-1.5ex]{Interface}
          & stacking layers       & length of Si-O (\AA) & polarization & (eV)           \\\hline\hline
I         & Mg-O-Si-Ni-Mn ($\gets$) & 1.86                 & 100\%        &                \\
II        & Mg-O-Si-vac-Mn ($\to$)  & 1.25             & 100\% & \raisebox{1.5ex}[-1.5ex]{54.5} \\ \hline
\end{tabular}
\label{details-111}
\end{table}

\begin{table}
\caption{Spin magnetic moments (per atom, in $\mu_B$) at the (111) interfaces
for the first three layers of NiMnSi and the first
two layers of MgO at the interface. Single and double
asterisks denote atoms in the sub-interface and the sub-sub-interface layers, respectively.}
\smallskip
\begin{tabular}{l||c|c|c|c|c|c}\hline
             & Ni*  & Mn** &    Si   & O    & Mg*  & total \\\hline\hline
Interface I  & 0.21 & 3.02 & $-$0.07 & 0.01 & 0.00 & 3.17 \\
Interface II & ---  & 3.27 & $-$0.06 & 0.00 & 0.00 & 3.21 \\ \hline
Bulk-like    & 0.13 & 3.06 & $-$0.11 & 0.00 & 0.00 & 3.07 \\ \hline
\end{tabular}
\label{moments-111}
\end{table}

Details of the NiMnSi/MgO (111) interface are provided in Table~\ref{details-111}. Since
the two interfaces, I and II, differ slightly, it is not possible to extract $W_{\rm I}$ and
$W_{\rm II}$ individually by computing the energy difference analogous to Eq.\ (1), and dividing
by two. Thus, for definiteness, we present only the aggregate work of separation,
$W_{\rm I+II} = 54.5$ eV. Note. however, that both, I and II, have direct Si-O bonds
at the interface, and differ only in the second neighboring atomic layer to O (Ni for
interface I vs.\ vacancies for interface II). Hence we may assume that the individual works
of separation, $W_{\rm I}$ and $W_{\rm II}$, are not so different from each other. Together
with the large value obtained, we therefore conclude that both interfaces are energetically
stable.

The average spin magnetic moments per atom at the interface are given in Table~\ref{moments-111},
considering the first three layers of NiMnSi and the first two layers of MgO.
Atoms in the atomic layer furthest away from the interface show bulk-like moments. For
interface I the Mn moment in the sub-sub-interface layer is very close to the bulk value,
whereas the Ni and Si moments are considerably larger than their bulk values. The
total magnetic moment of interface I shows little difference from the bulk since Mn
carries most of it. For interface II the Mn moment increases by 0.2$\mu_B$ as
compared with interface I and the bulk value, due to the vacancy layer next to it,
which suppresses hybridization between the Mn and Ni $d$ orbitals.

After structural optimization of the (111) interface
a relatively large reconstruction, as compared with the (100) interface, is encountered
as a consequence of the low symmetry and the presence of two nonequivalent interfaces.
The atoms in the interface layers are shifted away from their initial positions (bulk
structure), where the average difference in the $c$-coordinates of the Si and O atoms
is 1.86~\AA\ and 1.25~\AA\ for interface I and II, respectively. This is much shorter
than the bond lengths at interfaces with 100\% spin polarization
discovered so far, such as NiMnSb/CdS (111) with a Sb-S bond length of 2.7~\AA\
(Ref.\ \onlinecite{wijs01}) and NiMnSb/InP (111) with a Sb-P bond
length of 2.61~\AA\ (Ref.\ \onlinecite{galanakis05b}).
The authors of these studies have suggested that the rather long bonds for the latter two
structures might be essential for the conservation of half-metallicity. Our results for
NiMnSi/MgO (111), however, indicate that long bond lengths are not a necessary condition.

It must be noted that half-metallicity only persists in NiMnSb/CdS and NiMnSb/InP when the
structural optimization is partial (no interface relaxation).
In fact, it has been shown in Refs.\ \onlinecite{wijs01,galanakis05b} that a full structural
optimization leads to considerable reconstruction between interface layers and
consequently a significant suppression of the spin polarization. In contrast, for the
NiMnSi/MgO (111) interface studied here a 100\% spin polarization is obtained after full
structural optimization, whereas a drastic polarization decrease is found for partial
relaxation, i.e., when only the in-plane atomic positions and the $c$-axis are
optimized. Thus, the NiMnSi/MgO (111) scenario is, unexpectedly, at variance with
the usual understanding of the loss of half-metallicity by disorder and interdiffusion
at the interface. Whether the scenario of the full structural optimization is closer to
reality depends, of course, on the actual experimental conditions.

\section{Conclusion}

We have investigated the electronic and magnetic properties of the (100) interface
between the Heusler alloy NiMnSb and the insulator MgO. A full structural optimization
has been performed, showing that MnSb/OMg termination leads to the most favorable interface.
We find that the half-metallic property of the bulk Heusler alloy is lost at the interface.
The spin polarization at the MnSb-terminated interface is $\sim 60$\%, whereas Ni termination
suppresses it drastically, due to strong oxidation of Ni. For the NiMnSb/MgO (111) interface
the spin polarization at the Fermi level is also lost, but the energy gap in the minority band
still exists. This opens up the possibility of shifting the Fermi level back by proper doping.

For the NiMnSi/MgO (111) interface, on the other hand, half-metallicity is maintained
at both interfaces I and II. Besides, short bonds between Si and O are formed. All
these results apply to the structures after full optimization, including in-plane and
out-of-plane relaxation. However, the spin polarization at the interface
significantly deteriorates when only the in-plane atomic coordinates and the
$c$-axis of the supercell are optimized. Our results thus suggest that it is
promising to investigate experimentally the growth of NiMnSi (111) on MgO (111)
substrates, to shed further light on the mechanisms relevant for the fate of half-metallicity
at interfaces. In addition, NiMnSi/MgO (111) could be another attractive heterostructure
for spintronics applications.

\begin{acknowledgments}
The authors acknowledge helpful discussions with Cosima Schuster during the initial stages
of this work, and thank the Deutsche Forschungsgemeinschaft for financial support
(through TRR 80).
\end{acknowledgments}

\end{document}